\documentclass[12pt]{article}

\newcommand{\be}{\begin{equation}}
\newcommand{\ee}{\end{equation}}
\newcommand{\bi}[1]{\vspace{-3mm} \bibitem{#1}}
\usepackage{epsfig,amsmath,amssymb,graphics,graphicx}

\begin{document}

\begin{center}

{\it International Journal of Modern Physics B 21 (2007) 955-967}
\vskip 3 mm

{\Large \bf Fokker-Planck Equation for Fractional Systems}
\vskip 3 mm

{\large \bf Vasily E. Tarasov }\\

\vskip 3mm

{\it Skobeltsyn Institute of Nuclear Physics, \\
Moscow State University, Moscow 119991, Russia } \\
E-mail: tarasov@theory.sinp.msu.ru
\end{center}

%\vskip 3 mm

\begin{abstract}
The normalization condition, average values 
and  reduced distribution functions can be
generalized by fractional integrals. 
The interpretation of the fractional analog of phase space as 
a space with noninteger dimension is discussed. 
A fractional (power) system is described by 
the fractional powers of coordinates and momenta.
These systems can be considered as non-Hamiltonian systems 
in the usual phase space.
The generalizations of the Bogoliubov equations 
are derived from the Liouville equation for fractional (power) systems.
Using these equations, the corresponding
Fokker-Planck equation is obtained. 
\end{abstract}

\noindent
Keywords: Fokker-Planck equation; non-Hamiltonian systems; 
fractal; fractional integral

%%%\noindent
%%%PACS:  05.20.-y; 05.45.Df; 45.10.Hj\\
%05.45.Df Fractals
%47.53.+n Fractals in fluid dynamics
%%%05.45.-a Nonlinear dynamics and nonlinear dynamical systems 
%%%05.20.-y Classical statistical mechanics
%%%05.20.Gg Classical ensemble theory
%%%05.20.Dd Kinetic theory
%%%45.10.Hj  Perturbation and fractional calculus methods

\section{Introduction}

Fractional integrals and derivatives  \cite{SKM} have many
applications in statistical mechanics and kinetics 
\cite{Zaslavsky1,Zaslavsky2}. 
The generalization of the Fokker-Planck equation \cite{Chaos2005} 
can be used to describe kinetics in the fractal media.
It is known that the Fokker-Planck equation can be derived
from the Liouville and Bogoliubov equations \cite{Is,RL,F}.
The Liouville equation is obtained from the
normalization condition and from the Hamilton equations.
The Bogoliubov equations can be derived from the
Liouville equation and from the definition of the average value.
In this paper, the generalized Fokker-Planck equation 
is obtained from the Liouville and Bogoliubov equations
for fractional (power) systems.
For this aim, we use fractional generalizations of thenormalization condition 
and the average values \cite{nonHam1,nonHam2,nonHam3,nonHam4}. 

In the paper, we suggest the physical interpretation 
of integrals of noninteger order. 
The fractional integral is considered as an
integral on the fractal or noninteger-dimensional space.
This interpretation is connected with the definition of 
noninteger dimension.
We prove that fractional integration  
can be used to describe processes and systems on fractal.
The physical values on fractals can be "averaged" and 
the distribution of the values on fractal can be replaced 
by some continuous distribution. 
To describe the distribution on the set with 
noninteger dimension, we use the fractional integrals.
The order of the integral is equal to the fractal Hausdorff dimension
of the set.
The consistent approach to describe the distribution on fractal 
is connected with the mathematical definition of the integrals
on fractals \cite{RLWQ,Zverev,Svozil,PS}. 
It was proved \cite{RLWQ} that integrals on net of fractals 
can be approximated by fractional integrals. 
In Ref. \cite{nonHam1,nonHam2,nonHam3,nonHam4}, we proved that fractional 
integrals can be considered as integrals over the space with noninteger 
dimension up to a numerical factor. We use the well-known 
formulas of dimensional regularizations \cite{Col}.  
There is an interpretation that follows from the fractional measure
of phase space \cite{nonHam1,nonHam2,nonHam3,nonHam4}, 
which is used in the fractional integrals.
The fractional phase space can be considered as a space 
that is described by the fractional powers of coordinates and momenta. 
Using this phase space, we can consider some of the 
non-Hamiltonian systems as generalized Hamiltonian systems \cite{nonHam1,nonHam2,nonHam3,nonHam4}. 
The fractional systems can be described as 
exitations of the fractal medium \cite{nonHam1,nonHam2,nonHam3,nonHam4}. 

In Sec. 2, we consider the Hausdorff measure, the Hausdorff dimension 
and the integration on fractals to fix notation, and 
provide a convenient reference.
The connections of the integration on fractals and the fractional 
integrals are discussed.
The fractional average values and reduce distribution functions are defined.
In Sec. 3, we derive Fokker-Planck equations from the 
Liouville equation for fractional (power) systems.
A short conclusion is given in Sec. 4.

\section{Integration on fractal and fractional integration} 

\subsection{Hausdorff measure and Hausdorff dimension}

Fractals are measurable metric sets with a noninteger Hausdorff dimension.  
To define the Hausdorff measure and the Hausdorff dimension, we 
consider a measurable metric set $(W, \mu_H)$ with $W \subset \mathbb{R}^n$. 
The elements of $W$ are denoted by $x, y, z, . . . $, and represented by 
$n$-tuples of real numbers $x = (x_1,x_2,...,x_n)$ 
such that $W$ is embedded in $\mathbb{R}^n$. 
The set $W$ is restricted by the following conditions:
(1) $W$ is closed;
(2) $W$ is unbounded; 
(3) $W$ is regular (homogeneous, uniform) with its points randomly distributed.

The diameter of a subset $E \subset W \subset \mathbb{R}^n$ is 
\[ diam(E)=\sup\{ G(x,y): \ x , y  \in E \}  , \]
where $G(x,y)$ is a metric function of two points $x$ and $y \in W$. 

Let us consider a set $\{E_i\}$ of subsets $E_i$ such that
$diam(E_i) < \varepsilon$  $\forall i$, and 
$W \subset \bigcup^{\infty}_{i=1} E_i$.
Then, we define 
\be
\xi(E_i,D)= \omega(D) [diam (E_i)]^D .
\ee
The factor $\omega(D)$ depends on the geometry of $E_i$.
If $\{E_i\}$ is the set of all (closed or open) balls in $W$, then
\be
\omega(D)=\frac{\pi^{D/2} 2^{-D}}{\Gamma(D/2+1)}. 
\ee

The Hausdorff dimension $D$ of a subset $E \subset W$ 
is defined \cite{Federer} by 
\be \label{Hd}
D= dim_H (E)=\sup \{ d \in \mathbb{R}: \  \mu_H (E, d ) = \infty  \} 
=\inf\{ d \in \mathbb{R} : \ \mu_H(E,d)=0 \}.
\ee
From the definition (\ref{Hd}), we obtain \\
(1) $\mu_H(E,d)=0$ for $d>D=dim_H(E)$; \\ 
(2) $\mu_H(E,d)=\infty$ for $d<D=dim_H(E)$. 

The Hausdorff measure $\mu_H$ of a subset $E \subset W $ 
is defined \cite{Federer,R} by
\be 
\mu_H(E,D)= \omega(D) \lim_{diam(E_i) \rightarrow 0} 
\inf_{\{E_i\}} \sum^{\infty}_{i=1} [diam(E_i)]^D .
\ee
Note that $\mu_H(\lambda E,D) =\lambda^D \mu_H(E,D)$, 
where $\lambda >0$, and $\lambda E =\{ \lambda x, \ x \in E \}$.

%%%%%%%%%%%%%%%%%%%%%%%%%
\subsection{Function and integrals on fractal}

Let us consider the functions on $W$: 
\be \label{f}
f(x)=\sum^{\infty}_{i=1} \beta_i \chi_{E_i}(x) , 
\ee
where $\chi_{E}$ is the characteristic function of $E$: 
$\chi_{E}(x)=1$ if $x \in E$, and $\chi_{E}(x)=0$ if $x \not \in E$. 

The Lebesgue-Stieltjes integral for (\ref{f}) is defined by
\be \label{LSI}
\int_W f d \mu =\sum^{\infty}_{i=1} \beta_i \mu_H(E_i).
\ee
Therefore
\be \label{int}
\int_W f(x) d \mu_H (x) =
\lim_{ diam(E_i) \rightarrow 0} \sum_{E_i} f(x_i) \xi(E_i,D)= 
\omega(D) \lim_{ diam(E_i) \rightarrow 0} \sum_{E_i} f(x_i) [diam(E_i)]^D .
\ee
It is possible to divide $\mathbb{R}^n$ into parallelepipeds 
\be
E_{i_1...i_n} =\{  (x_1,...,x_n) \in W: \quad x_j =
(i_j-1) \Delta x_j +\alpha_j, \quad
0 \le \alpha_j \le \Delta x_j, \quad j=1,...,n \} .
\ee
Then
\be
d \mu_H (x)= \lim_{diam(E_{i_1...i_n}) \rightarrow 0}
\xi(E_{i_1 ... i_n},D)=
\lim_{diam(E_{i_1 ... i_n}) \rightarrow 0}
\prod^n_{j=1} (\Delta x_j)^{D/n}=\prod^n_{j=1} d^{D/n} x_j .
\ee
The range of integration $W$ can be parametrized by 
polar coordinates with $r=G(x, 0)$ and angle $\Omega$. 
Then $E_{r,\Omega}$ can be thought of as a spherically 
symmetric covering around a center at the origin. 
In the limit, function $\xi(E_{r,\Omega},D)$ gives
\be \label{rrr}
d\mu_H(r,\Omega)=\lim_{diam(E_{r, \Omega}) \rightarrow 0}
\xi (E_{r,\Omega},D)=d\Omega^{D-1} r^{D-1} dr. 
\ee

Let us consider $f(x)$ that is symmetric with respect 
to some point $x_0 \in W$, 
i.e. $f(x) = const$ for all $x$ such that $G(x, x_0)=r$ 
for arbitrary values of $r$. Then the transformation
\be \label{WrZ}
W \rightarrow W^{\prime} : \ x \rightarrow x^{\prime}=x-x_0
\ee
can be performed to shift the center of symmetry.  
Since $W$ is not a linear space, 
(\ref{WrZ}) need not be a map of $W$ onto itself. 
Map (\ref{WrZ}) is measure preserving. 
Using (\ref{rrr}),
the  integral over a $D$-dimensional metric space is  defined by
\be \label{intWf}
\int_W f d\mu_H = \frac{2 \pi^{D/2}}{\Gamma(D/2)}  
\int^{\infty}_0 f(r) r^{D-1} dr .
\ee
This integral is known in the theory of the 
fractional calculus \cite{SKM}. 
The right Riemann-Liouville fractional integral is
\be \label{FID}
I^{D}_{-} f(z)=\frac{1}{\Gamma(D)} \int^{\infty}_z (x-z)^{D-1} f(x) dx .
\ee
Equation (\ref{intWf}) is reproduced by
\be \label{FIFI}
\int_W f d\mu_H = \frac{2 \pi^{D/2} \Gamma(D)}{\Gamma(D/2)} I^{D}_{-} f(0) .
\ee
Relation (\ref{FIFI}) connects the integral on fractal 
with the integral of fractional order.
This result permits to apply different tools of the fractional calculus
\cite{SKM} for the fractal medium.
As a result, the fractional integral can be considered as an
integral on fractal up to the numerical factor 
$\Gamma(D/2) /[ 2 \pi^{D/2} \Gamma(D)]$.

Note that the interpretation of fractional integral
is connected with the fractional dimension 
\cite{nonHam1,nonHam2,nonHam3,nonHam4}.
This interpretation follows from the well-known formulas 
for dimensional regularization \cite{Col}.
The fractional integral can be considered as an
integral in the noninteger-dimensional space
up to the numerical factor $\Gamma(D/2) /[ 2 \pi^{D/2} \Gamma(D)]$.
It was proved \cite{Svozil} that the fractal space-time 
approach is technically identical with the dimensional regularization.

The integral defined in (\ref{int}) satisfies 
the following properties:

(1) Linearity:
\be \label{14}
\int_X (af_1+bf_2) d \mu_H=a \int_X f_1 d \mu_H+b \int_X f_2 d \mu_H ,
\ee
where $f_1$ and $f_2$ are arbitrary functions; 
$a$ and $b$ are arbitrary constants.

(2) Translational invariance:
\be \label{15}
\int_X f(x+x_0) d \mu_H(x)= \int_X f(x) d \mu_H(x)
\ee
since $d \mu_H(x - x_0)=d \mu_H(x)$ as a consequence 
of homogeneity (uniformity).

(3) Scaling property:
\be \label{16}
\int_X f(ax) d \mu_H(x)= a^{-D}\int_X f(x) d \mu_H(x)
\ee
since $d \mu_H (x/a)=a^{-D}d \mu_H(x)$.

\noindent
It has been shown \cite{Col} that conditions (\ref{14})-(\ref{16}) 
define the integral up to normalization \cite{Col}.

\subsection{Multi-variable integration on fractal}

The integral (\ref{intWf}) is defined for a single variable, 
and not multiple variables. 
It is useful for integrating spherically symmetric functions. 
This integral can be generalized for the multiple variables 
by using the product spaces and product measures.

Let us consider the measure spaces $(W_k,\mu_k,D)$ with $k=1,2,3$,
and form a Cartesian product of the sets $W_k$ producing the space 
$W=W_1 \times W_2 \times W_3$.
The definition of product measures and 
the application of the Fubini's theorem 
provides a measure for $W$ as
\be
(\mu_1 \times \mu_2 \times \mu_3)(W) 
= \mu_1(W_1) \mu_2(W_2) \mu_3(W_3). 
\ee
The integration over a function $f$ on the product space is
\be \label{int-n}
\int f ({\bf r}) \; d \mu_1 \times \mu_2 \times \mu_3 = 
\int \int \int f (x_1,x_2 , x_3) \; d \mu_1(x_1) d \mu_2(x_2) d \mu_3(x_3). 
\ee
In this form, the single-variable measure from (\ref{intWf}) may 
be used for each coordinate $x_k$, which has 
an associated dimension $\alpha_k$:
\be 
d \mu_k(x_k) = \frac{2 \pi^{\alpha_k/2}}{\Gamma(\alpha_k/2)} 
|x_k|^{\alpha_k-1} dx_k , \quad k=1,2,3.
\ee
The total dimension of  $W=W_1 \times W_2 \times W_3$ is  
$D= \alpha_1+\alpha_2+\alpha_3$. 

Let us reproduce the result (\ref{intWf}) from (\ref{int-n}).  
We take a spherically symmetric function 
$f({\bf r})=f (x_1, x_2 , x_3) = f (r)$,
where $r^2 = (x_1)^2 + (x_2)^2 + (x_3)^2$.
Equation (\ref{int-n}) becomes
\[
\int d \mu_1(x_1) d\mu_2(x_2) d\mu_3(x_3) f (x_1,x_2,x_3) = 
\]
\[
=\frac{2 \pi^{\alpha_1/2}}{\Gamma(\alpha_1/2)}
\frac{2 \pi^{\alpha_2/2}}{\Gamma(\alpha_2/2)}
\frac{2 \pi^{\alpha_3/2}}{\Gamma(\alpha_3/2)}
\int dr \int d \phi \int d \theta \; 
J_3(r,\phi) \; r^{ \alpha_1+\alpha_2+ \alpha_3- 3} \times
\]
\be \label{21}
\times (\cos \phi)^{\alpha_1-1} (\sin \phi)^{\alpha_2+ \alpha_3 -2} 
(\sin \theta)^{\alpha_3 -1}  f(r) ,
\ee
where $J_3(r,\phi)= r^2 \sin \phi$ is the Jacobian of the coordinate change.

To perform the integration in spherical 
coordinates $(r, \phi, \theta)$, we use
\be
\int^{\pi/2}_0 \sin^{\mu-1} x \cos^{\nu-1} x dx =
\frac{\Gamma(\mu/2) \Gamma(\nu/2)}{2 \Gamma([\mu+\nu]/2)} ,
\ee
where $\mu >0$, $\nu>0$. Then Eq. (\ref{21}) becomes
\be
\int d\mu_1(x_1) d \mu_2(x_2) d\mu_3(x_3)f (r) = 
\frac{2 \pi^{D/2}}{\Gamma(D/2)} \int f(r) r^{D-1} dr .
\ee
This equation describes integration over 
a spherically symmetric function 
in $D$-dimensional space and 
reproduces result (\ref{intWf}).

\subsection{Probability distribution on fractal}

The probability that is distributed in the three-dimensional 
Euclidean space is defined by
\be \label{MW} P_3(W)=\int_W \rho({\bf r}) d V_3 , \ee
where $\rho({\bf r})$ is the density of probability distribution,
and $dV_3=dx dydz$ for the Cartesian coordinates.
%%%The fractional integrals can be used to describe fields that 
%%%are defined on the set $W$ with fractional Hausdorff dimension $dim_H(W)=D$.

If we consider the probability that is distributed on the 
measurable metric set $W$ with the fractional Hausdorff dimension $D$,
then the probability is defined by the integral 
\be \label{35}
P_D(W)= \int_W \rho({\bf r}) dV_D, 
\ee 
where $D =dim_H (W)=\alpha_1+\alpha_2+\alpha_3$, and 
\be
dV_D=d \mu_1(x_1) d \mu_2(x_2) d \mu_3(x_3)=c_3(D,{\bf r}) dV_3, ,
\ee
\be
c_3(D,{\bf r})=\frac{8 \pi^{D/2}}{\Gamma(\alpha_1) 
\Gamma(\alpha_2) \Gamma(\alpha_3) }
|x|^{\alpha_1-1} |y|^{\alpha_2-1} |z|^{\alpha_3-1}  .
\ee
There are many different definitions of fractional integrals \cite{SKM}.
For the Riemann-Liouville fractional integral, 
the function $c_3(D,{\bf r})$ is 
\be \label{c3Dr}
c_3(D,{\bf r})=\frac{ |x|^{\alpha_1-1}|y|^{\alpha_2-1} |z|^{\alpha_3-1} }{
\Gamma(\alpha_1) \Gamma(\alpha_2) \Gamma(\alpha_3) },
\ee
where $x$, $y$, $z$ are the Cartesian coordinates, 
and $D=\alpha_1+\alpha_2+\alpha_3$, \ $0<D\le 3$.
As the result, we obtain the Riemann-Liouville fractional 
integral \cite{SKM} in Eq. (\ref{35}) up to numerical factor $8 \pi^{D/2}$.
Therefore, Eq. (\ref{35}) can be considered as a
fractional generalization of Eq. (\ref{MW}). 

For $\rho({\bf r})=\rho(|{\bf r}|)$, we can use
the Riesz definition of the fractional integrals \cite{SKM}.
Then  
\be \label{IDc}
c_3(D,{\bf r})=\frac{\Gamma(1/2)}{2^D \pi^{3/2} \Gamma(D/2)} |{\bf r}|^{D-3} , 
\ee
Note that
\be
\lim_{D\rightarrow 3-} c_3(D,{\bf r}) = (4 \pi^{3/2})^{-1} .
\ee
Therefore, we suggest to use 
\be \label{lam2}
c_3(D,{\bf r})=\frac{2^{3-D} \Gamma(3/2)}{\Gamma(D/2)} |{\bf r}|^{D-3} .
\ee
Definition (\ref{lam2}) allows us to derive the usual integral 
in the limit $D\rightarrow (3-0)$.

For $D=2$, Eq. (\ref{35}) gives the fractal probability 
distribution in the volume. 
In general, it is not equivalent to the distribution
on the two-dimensional surface.
Equation (\ref{c3Dr}) is equal (up to numerical factor $8 \pi^{D/2}$) 
to the integral on the measurable metric set $W$ with Hausdorff 
dimension $dim_H(W)=D$.
To have the usual dimensions of the physical values,
we can use vector ${\bf r}$, and coordinates 
$x$, $y$, $z$ as dimensionless values.

\subsection{Fractional average values}

To derive the fractional analog of the average value, 
we consider the fractional integral for function $f(x)$.
If function $f(x)$ is equal to the distribution function
$\rho(x)$, then we can derive the normalization condition.
If the function $f(x)$ is equal to the multiplication of
distribution function $\rho(x)$ and classical observable
$A(x)$, then we have the definition of the fractional average value.

The fractional generalization of the average value \cite{nonHam1,nonHam2,nonHam3,nonHam4} 
can be present by
\be \label{Aa} <A>_{\alpha}=
(I^{\alpha}_{+}A\rho)(y)+(I^{\alpha}_{-}A\rho)(y) , \ee
where
\be  \label{I+}I^{\alpha}_{+}f=
\frac{1}{\Gamma (\alpha)} \int^{y}_{-\infty}
\frac{f(x)dx}{(y-x)^{1-\alpha}} , \quad
 I^{\alpha}_{-}f=
\frac{1}{\Gamma (\alpha)} \int^{\infty}_{y}
\frac{f(x)dx}{(x-y)^{1-\alpha}} . \ee
For $\alpha=1$, Eq. (\ref{Aa})  gives the usual average value.

The fractional average value (\ref{Aa}) can be written \cite{nonHam1,nonHam2,nonHam3,nonHam4} as
\be \label{FI5} <A>_{\alpha}= \frac{1}{2}
\int^{\infty}_{-\infty} [ (A\rho)(y-x)+ (A\rho)(y+x) ] d\mu_{\alpha}(x) , \ee
where 
\be \label{dm}
d\mu_{\alpha}(x)=\frac{|x|^{\alpha-1} dx}{\Gamma(\alpha)}=
\frac{d x^{\alpha}}{\alpha \Gamma(\alpha)} .\ee
Here, we use 
\be \label{xa}
x^{\alpha} = sgn(x) |x|^{\alpha} , \ee
where the function $sgn(x)$ is equal to $+1$ for $x\ge0$,
and $-1$ for $x<0$.

%%%%%%%%%%%%%%%%%%%%%%%%%%%%%%%%%%%%%%%%%%%%%%%%%%%%%%%%%%%%%%%%%%%%%%%

Let us introduce notations to consider the fractional
average value for phase space.

(1) The operator $T_{x_k}$ is defined by
\be \label{T} T_{x_k} f(...,x_k,...)= 
\frac{1}{2}\Bigl( f(...,x^{\prime}_k-x_k,...)
+f(...,x^{\prime}_k+x_k,...) \Bigr) . \ee
For $k$-particle, which is described by
coordinates $q_{ks}$ and momenta $p_{ks}$, ($s=1,...,m$),
the operator $T[k]$ is 
\be T[k]=T_{q_{k1}} T_{p_{k1}}...T_{q_{km}} T_{p_{km}} . \ee
For the $n$-particle system, we define the operator
$T[1,...,n]=T[1]...T[n]$. 

(2) The operator $\hat I^{\alpha}_{x_k}$ is defined by
\be \hat I^{\alpha}_{x_k} f(x_k)=
\int^{+\infty}_{-\infty}  f(x_k) d \mu_{\alpha} (x_k) .\ee
Then the fractional integral (\ref{FI5}) can be rewritten in the form
\[ <A>_{\alpha}=\hat I^{\alpha}_{x} T_x A(x)\rho(x) . \]
The integral operator 
$\hat I^{\alpha}[k]=\hat I^{\alpha}_{q_{k1}} \hat I^{\alpha}_{p_{k1}} ...
\hat I^{\alpha}_{q_{km}} \hat I^{\alpha}_{p_{km}}$ is
\be \hat I^{\alpha}[k] f({\bf q}_k,{\bf p}_k)=
\int f({\bf q}_k,{\bf p}_k)
d \mu_{\alpha}({\bf q}_k,{\bf p}_k) ,  \ee
where 
\[ d \mu_{\alpha}({\bf q}_k,{\bf p}_k)=(\alpha \Gamma(\alpha))^{-2m}
d q^{\alpha}_{k1}\wedge dp^{\alpha}_{k1} \wedge ... \wedge
d q^{\alpha}_{km}\wedge dp^{\alpha}_{km} . \]
For the $n$-particle system, we use 
$\hat I^{\alpha}[1,...,n]=\hat I^{\alpha}[1]...\hat I^{\alpha}[n]$. 

The fractional average values for the $n$-particle system 
is defined \cite{nonHam1,nonHam2,nonHam3,nonHam4} by
\be <A>_{\alpha}=
\hat I^{\alpha}[1,...,n] T[1,...,n] A \rho_{n} .  \ee
In the simple case ($n=m=1$), the fractional average value is 
\be \label{AV2} <A>_{\alpha}=
\int^{+\infty}_{-\infty} \int^{+\infty}_{-\infty}
d\mu_{\alpha}(q,p) 
\ T_q T_p A(q,p)\rho(q,p) . \ee
Note that the fractional normalization 
condition \cite{nonHam1,nonHam2,nonHam3,nonHam4} is a special case of 
this definition of average values $<1>_{\alpha}=1$.

\newpage
%%%%%%%%%%%%%%%%%%%%%%%%%%%%%%%%%%%%%%%%%%%%%%%%%%%%%%%%%%%%%%%%%%%%
\section{Fokker-Planck equation from Liouville equation}

Let us consider a system with $n$ identical particles
and the Brownian particle.
The distribution function of this system is denoted by
$\rho_{n+1}({\bf q},{\bf p},Q,P,t)$, where
\[ {\bf q}=({\bf q}_1,...,{\bf q}_n), \quad {\bf q}_k=(q_{k1},...,q_{km}) , \]
\[ {\bf p}=({\bf p}_1,...,{\bf p}_n), \quad {\bf p}_k=(p_{k1},...,p_{km})  \]
are coordinates and momenta of the particles; 
$Q=(Q_s)$ and $P=(P_s)$ ($s=1,..,m$) 
are coordinates and momenta of Brownian particles.
The fractional normalization condition \cite{nonHam1,nonHam2,nonHam3,nonHam4} has the form
\be \label{nc} \hat I^{\alpha}[1,...,n,n+1] \tilde \rho_{n+1}=1 , \ee
where
\be \tilde \rho_{n+1}=T[1,...,n,n+1] \rho_{n+1}({\bf q},{\bf p},Q,P,t) .\ee
The reduced distribution function for the Brownian particle is
\be \label{rB} \tilde \rho_{B}(Q,P,t)=
\hat I^{\alpha}[1,...,n] \tilde \rho_{n+1}({\bf q},{\bf p},Q,P,t) . \ee

The distribution $\tilde \rho_{n+1}$ satisfies the
Liouville equation \cite{nonHam1,nonHam2,nonHam3,nonHam4}:
\be \frac{\partial \tilde \rho_{n+1}}{\partial t}-
i(L_n+L_B)\tilde \rho_{n+1}=0 ,\ee
where $L_n$ and $L_B$ are Liouville operators such that
\be -iL_n\rho = \sum^{n,m}_{k,s} \left(
\frac{\partial (G^k_s \rho)}{\partial q^{\alpha}_{ks}} +
\frac{\partial (F^k_s \rho)}{\partial p^{\alpha}_{ks}}  \right), \ee
\be -iL_B\rho = \sum^{n,m}_{k,s} \left(
\frac{\partial (g_s \rho)}{\partial Q^{\alpha}_{s}} +
\frac{\partial (f_s \rho)}{\partial P^{\alpha}_{s}}  \right). \ee

The forces $F^k_s$ and $f_s$, and the velocities  $G^k_s$ and $g_s$
are defined by the Hamilton equations of motion.
For $k$th particle,
\be \label{HE1}\frac{d q^{\alpha}_{ks}}{dt}=G^k_s({\bf q},{\bf p}), \quad
\frac{d p^{\alpha}_{ks}}{dt}=F^k_s({\bf q},{\bf p},Q,P) , \ee
where $k=1,...,n$. The Hamilton equations for the Brownian particle are
\be \label{HE2} \frac{d Q^{\alpha}_{s}}{dt}=g_s(Q,P), \quad
\frac{d P^{\alpha}_{s}}{dt}=f_s({\bf q},{\bf p},Q,P) . \ee
For simplification, we suppose
\be \label{Gg} G^k_s=p^{\alpha}_{ks}/m , \quad g_s=P^{\alpha}_s/M , \ee
where $M \gg m$.

Let us use the boundary condition in the form 
\be \label{bound} \lim_{t \rightarrow - \infty}
\rho_{n+1} ({\bf q},{\bf p},Q,P,t) =
\rho_n({\bf q},{\bf p},Q,T) \rho_B(Q,P,t) , \ee
where $\rho_n$ is a canonical Gibbs distribution function
\be \label{Gibbs} \rho_n({\bf q},{\bf p},Q,T)=
\exp \, \beta \Bigl[ {\cal F}-H({\bf q},{\bf p},Q) \Bigr] . \ee
Here, $H ({\bf q},{\bf p},Q)$ is a Hamilton function
\be H ({\bf q},{\bf p},Q)=
H_n({\bf q},{\bf p})+\sum^n_{k=1} U_B({\bf q}_k,Q) , \ee
where $H_n$ is a Hamiltonian of the $n$-particle system, and $U_B$
is an energy of interaction between particles and the Brownian particle.
If we use Eqs. (\ref{HE1}) and (\ref{Gg}), then
\be  H_n({\bf q},{\bf p})=\sum^{n,m}_{k,s} \frac{p^{2\alpha}}{2m}+
\sum_{k<l}U({\bf q}_k,{\bf q}_l) . \ee

The boundary condition (\ref{bound}) can be realized \cite{Zub} by the
infinitesimal source term in the Liouville equation
\be \label{epsl}
\frac{\partial \tilde \rho_{n+1}}{\partial t}-i(L_n+L_B)\tilde \rho_{n+1}=
- \varepsilon (\tilde \rho_{n+1}-\tilde \rho_n \tilde \rho_B) . \ee
Integrating (\ref{epsl}) by $\hat I^{\alpha}[1,...,n]$,
we obtain the equation for the Brownian particle distribution
\be \label{rhoB}
\frac{\partial \tilde \rho_B}{\partial t}
+\sum^{m}_{s=1} \frac{\partial (g_s \tilde \rho_B)}{\partial Q^{\alpha}_s}
+\hat I^{\alpha}[1,...,n]
\sum^m_{s=1}\frac{\partial (f_s \rho_{n+1})}{\partial P^{\alpha}_{s}}=0 . \ee
The formal solution \cite{Zub} has the form
\be \tilde \rho_{n+1}(t)=\varepsilon \int^0_{-\infty}
d \tau \ e^{\varepsilon \tau}
e^{-i \tau (L_n+L_B)} \tilde \rho_B (t+\tau) \tilde \rho_n , \ee
or
\be \label{88}
\tilde \rho_{n+1}(t)=\tilde \rho_B(t) \tilde \rho_n
- \int^0_{-\infty} d \tau \ e^{\varepsilon \tau}
e^{-i \tau (L_n+L_B)} 
\left( \frac{\partial}{\partial \tau}-i(L_n+L_B) \right)
\tilde \rho_B (t+\tau) \tilde \rho_n . \ee
Substituting (\ref{88}) into (\ref{rhoB}), we get
\[ \frac{\partial \tilde \rho_B}{\partial t}
+\sum^{m}_{s=1} \frac{\partial (g_s \tilde \rho_B)}{\partial Q^{\alpha}_s}
+\sum^m_{s=1}\frac{\partial \rho_B}{\partial P^{\alpha}_s}
\hat I^{\alpha}[1,...,n] (f_s \tilde \rho_n)- \]
\be \label{rhoB2}
-\hat I^{\alpha}[1,...,n]\sum^m_{s=1}
\frac{\partial}{\partial P^{\alpha}_{s}}
\int^0_{-\infty} d \tau \ e^{\varepsilon \tau}
e^{-i \tau (L_n+L_B)} 
\left( \frac{\partial}{\partial \tau}-i(L_n+L_B) \right)
\tilde \rho_B (t+\tau) \tilde \rho_n=0 . \ee
Note that
$\hat I^{\alpha}[1,..,n] f_s \tilde \rho_n$
can be considered as an average value of the force $f_s$.
This average value for the canonical Gibbs distribution (\ref{Gibbs})
is equal to zero. The last term can be simplified.
Using
\be  \frac{\partial \rho_n}{\partial Q^{\alpha}_s}=
\frac{1}{kT} f^{(p)}_s \rho_n , \ee
where $f^{(p)}_s$ is a potential force
\be \label{fp}
f^{(p)}_s=- \frac{\partial U_B}{\partial Q^{\alpha}_s}, \ee
we get
\[ -iL_B \tilde \rho_{n+1}=
\left( \frac{P_sf^{(p)t}_s}{MkT} \rho_B +
\frac{\partial (g_s \tilde \rho_B)}{\partial Q^{\alpha}_s}+
\frac{\partial (f_s \tilde \rho_B)}{\partial P^{\alpha}_s} \right) \rho_n  . \]

It can be proved by interactions that the term
\be \frac{\partial \tilde \rho_B}{\partial t}+
\frac{\partial (g_s \tilde \rho_B)}{\partial Q^{\alpha}_s} \ee
in the integral of (\ref{rhoB2}) does not contribute.
Then
\[ \frac{\partial \tilde \rho_B}{\partial t}
+\sum^{m}_{s=1} \frac{\partial (g_s \tilde \rho_B)}{\partial Q^{\alpha}_s}
+\sum^m_{s=1}\frac{\partial}{\partial P^{\alpha}_s}
\hat I^{\alpha}[1,...,n] \int^0_{-\infty} d \tau \ e^{\varepsilon \tau}
 f_s e^{-i \tau (L_n+L_B)} \tilde \rho_n  \cdot \]
\be \label{rhoB3} \cdot
\left( \frac{\partial(f_{s'}
\tilde \rho_B(t+\tau))}{\partial P^{\alpha}_{s'}}
+\beta M^{-1} f_{s'} P_{s'} \tilde \rho_B (t+\tau) \right)=0 . \ee
This equation is a closed integro-differential equation for 
the reduced distribution function $\tilde \rho_B$.
Note that force $f_s$ can be presented in the form
\[ f_s=f^{(p)}_s+f^{(n)}_s ,\]
where $f^{(p)}_s$ is a potential force (\ref{fp}), and
$f^{(n)}_s$ is a non-potential force that acts on the Brownian particle.
For the equilibrium approximation
$P \sim (MkT)^{1/2}$, $iL_B \sim M^{-1/2}$,
$iL_n \sim m^{-1/2}$, and $M \gg m$, we can use perturbation theory.

Using the approximation $\rho_B(t+\tau)=\rho_B(t)$
for Eq. (\ref{rhoB3}), we obtain the Fokker-Planck equation
for fractional power systems
%%$m<<M$ ($L_B<<L_n$)
\be \label{rhoB4} 
\frac{\partial \tilde \rho_B}{\partial t}
+\sum^{m}_{s=1} \frac{\partial (g_s\tilde \rho_B)}{\partial Q^{\alpha}_s}
+\sum^m_{s=1}\frac{\partial}{\partial P^{\alpha}_s}
\left( \frac{M}{\beta}
\frac{\partial( \gamma^1_{ss'} \tilde \rho_B(t))}{\partial P^{\alpha}_{s'}}
+ \gamma^2_{ss'} P_{s'} \tilde \rho_B (t) \right)=0 , \ee
where
\be \label{gamma1}
\gamma^1_{ss'}=\beta M \hat I^{\alpha}[1,...,n]
\int^0_{-\infty} d \tau \ e^{\varepsilon \tau}
f_s e^{-i \tau L_n} f_{s' }\tilde \rho_n . \ee
\be \label{gamma2}
\gamma^2_{ss'}=\beta M \hat I^{\alpha}[1,...,n]
\int^0_{-\infty} d \tau \ e^{\varepsilon \tau}
f_s e^{-i \tau L_n} f^{(p)}_{s' } \tilde \rho_n . \ee
If $f_s=f^{(p)}_s$, then $\gamma^1_{ss'}=\gamma^2_{ss'}$.

%%%%%%%%%%%%%%%%%%%%%%%%%%%%%%%%%%%%%%%%%%%%%%%%%%%%%%%%%

Let us consider the one-dimensional stationary Fokker-Planck 
equation (\ref{rhoB4}) with 
\[ {\partial (g_s\tilde \rho_B)}/{\partial Q^{\alpha}_s}=0 . \]
Then
\be \label{sFP1} 
\frac{\partial}{\partial P^{\alpha}}
\left( \frac{M}{\beta}
\frac{\partial( \gamma^1(P) \tilde \rho_B(t))}{\partial P^{\alpha}}
+ \gamma^2(P) P \tilde \rho_B (t) \right)=0 . \ee
Obviously, we get the relation 
\be \label{sFP2} 
\frac{M}{\beta}
\frac{\partial( \gamma^1 (P) \tilde \rho_B(t))}{\partial P^{\alpha}}
+ \gamma^2 (P) P \tilde \rho_B (t)=const . \ee
Assuming that the constant is equal to zero, we get
\be \label{sFP3} 
\frac{\partial [ \gamma^1 (P) \tilde \rho_B(t)]}{\partial P^{\alpha}}=
\frac{\beta \gamma^2 (P) P}{M} \tilde \rho_B (t) , \ee
or, in an equivalent form
\be \label{sFP4} 
\frac{\partial ln [\gamma^1 (P) \tilde \rho_B(t)] }{\partial P^{\alpha}}=
\frac{\beta \gamma^2 (P) P}{\gamma_1 (P)M} . \ee
The solution is
\be \label{sFP5} 
ln [\gamma^1 (P) \tilde \rho_B(t)]= 
\int  \frac{\beta \gamma^2 (P) P}{M\gamma_1 (P)} dP^{\alpha} +const .
\ee
As the result, we obtain 
\be \label{sFP6}
\tilde \rho_B(t)=\frac{N}{\gamma^1 (P)} 
\int \frac{\beta \gamma^2 (P) P}{M \gamma_1 (P)} dP^{\alpha},
\ee
where the coefficient $N$ is defined by the normalization condition.
Equation (\ref{sFP6}) describes the solution of 
the stationary Fokker-Planck equation for the fractional (power) system.
The special cases of (\ref{sFP6}) can be derived as 
done in Ref. \cite{Chaos2005}.\\

%%%%%%%%%%%%%%%%%%%%%%%%%%%%%%%%%%%%%%%%%%%%%%%%%%%%%%%%%%%%%%%
\section{Conclusion}

In this paper, the fractional generalizations of the
average value and the reduced distribution functions are used. 
The generalization of the Liouville and Bogoliubov equations 
are derived \cite{nonHam1,nonHam2,nonHam3,nonHam4} 
from the fractional normalization condition.
Using these equations, we obtain the Fokker-Planck equation
for fractional (power) systems.

The Liouville, Bogoliubov and Vlasov equations 
for fractional systems \cite{nonHam1,nonHam2,nonHam3,nonHam4} 
can be considered as equations in the noninteger-dimensional phase space.
For example, the systems that live on some fractals 
can be described by these equations.
Note that the fractional systems can be presented as special
non-Hamiltonian systems \cite{nonHam1,nonHam2,nonHam3,nonHam4}.
Fractional oscillators can be interpreted 
as elementary excitations of some fractal medium 
with noninteger mass dimension \cite{nonHam1,nonHam2,nonHam3,nonHam4}.
The fractional (power) systems are connected 
with the non-Gaussian statistics.
Classical dissipative and non-Hamiltonian systems can have
the canonical Gibbs distribution as a solution of the stationary
equations \cite{NHSM1,NHSM2}.
Using the methods \cite{NHSM1,NHSM2}, it is easy to prove
that some fractional dissipative systems can have
the fractional analog of the Gibbs distribution (non-Gaussian statistic)
as a solution of the stationary equations for fractional systems. \\

%%%%%%%%%%%%%%%%%%%%%%%%%%%%%%%%%%%%%%%%%%%%%%

\end{document}